# Bias of the SIR filter in estimation of the state transition noise

Tiancheng Li

T. Li is with the School of Mechatronics, Northwestern Polytechnical University, Xi'an, 710072, China. He is also with the Center for Automated and Robotics NDT, London South Bank University, London, SE1 0AA, UK (Email: lit3@lsbu.ac.uk; robottcli@gmail.com; tiancheng.li1985@gmail.com)

*Abstract*

This Note investigates the bias of the sampling importance resampling (SIR) filter in estimation of the state transition noise in the state space model. The SIR filter may suffer from sample impoverishment that is caused by the resampling and therefore will benefit from a sampling proposal that has a heavier tail, e.g. the state transition noise simulated for particle preparation is bigger than the true noise involved with the state dynamics. This is because a comparably big transition noise used for particle propagation can spread overlapped particles to counteract impoverishment, giving better approximation of the posterior. As such, the SIR filter tends to yield a biased (bigger-than-the-truth) estimate of the transition noise if it is unknown and needs to be estimated, at least, in the forward-only filtering estimation. The bias is elaborated via the direct roughening approach by means of both qualitative logical deduction and quantitative numerical simulation.

---

## I. Introduction:

Nonlinear and non-Gaussian state space models (SSMs) widely exist in the field of control and single processing, often in which the observed data may be used to estimate the parameter(s) of the model (in the process of estimating the state) if it is unknown. The goal of parameter estimation is to compute an estimate of the true parameter that can best match the observations, being 'most likely' in a specific principle. This Note concerns with estimating the state transition noise by using the sampling importance resampling (SIR) filter, which is the most commonly used type of particle filters.

The SIR filter has been widely applied for parameter estimation since it relaxes linearity and Gaussian assumptions see e.g. [1, 2, 3, 4]. However, divergences/biases have been observed in experiments [1, 2, 3], which could not be ruled out completely. The bias was abstractly attributed to the sample degeneracy [3, 4], without adequate elaboration. The theory suggests that the particle filter benefits from a sampling proposal function that that has a heavier tail so that the filter is sensitive to outliers. This just indicates that a biased estimation of the state transition noise may occur. The bias is elaborated here with the sample impoverishment caused by resampling in the SIR filter. In what follows, qualitative logical deduction (section II) and quantitative simulation evidences (III) are presented for illustration.

## II. Qualitative logical deduction

### A. Estimation of the state transition noise

In the context of the general non-linear non-Gaussian SSM, if is often of great interest and significance to estimate the parameter(s) of the model in the process of estimating the state. The SSM can be written by two recursive equations

$$x_k = f_{k|k-1}(x_{k-1}, \theta_k) \quad \text{(state dynamic equation)} \tag{1}$$

$$y_k = g_{k|k}(x_k, \beta_k) \quad \text{(observation equation)} \tag{2}$$

where $k$ indicates time, $x_k$ denotes the state, $y_k$ denotes the observation, $\theta_k$ and $\beta_k$ denote noises affecting the state Markov transition equation $f_{k|k-1}(\cdot)$, and observation equation $g_{k|k}(\cdot)$, respectively and this Note focuses particularly on estimating the static transition noise $\theta_k$ by the SIR filter, i.e. for $t=1, 2,\ldots, k$, $\theta_t=\theta^*$, where $\theta^*$ is a static parameter. Let $y_{1:k} \triangleq (y_1, y_2, ..., y_k)$ be the history path of the observation process.

Assuming that true noise $\theta^*$ are involved with the state, generating observations $y_{1:k}$ and the values are unknown, our goal is to compute point estimates of $\theta^*$ from the observations. In general maximum likelihood (ML) principles, the estimate of $\theta^*$ is the maximizing argument of the marginal likelihood of the observed data

$$\hat{\theta} = \arg\max_{\theta \in \Omega} p(y_{1:k} | \theta) \tag{3}$$

where $\Omega$ is a specified parameter space.

**Remark 1.** $\forall$ two parameters $\theta_1$ and $\theta_2$ that are close-enough to each other in the monotonic domain of Eq. (3): if using transition noise $\theta_1$ for particle propagation obtains better approximation of the posterior in the SIR filter than using $\theta_2$, i.e. parameters $\theta_1$ is more likely than $\theta_2$ to generate the observations,

$$p(y_{1:k} | \theta_1) > p(y_{1:k} | \theta_2) \tag{4}$$

then, the SIR filter will more likely obtain an estimate of the true parameter $\theta^*$ that is closer to $\theta_1$ than $\theta_2$.

**Proof.** This Remark is just the content of the ML principle. By saying a better approximation of the posterior, it means that the underlying particle approximation of the state is closer to the real state and is therefore more likely to match the observation. $\theta_1$ and $\theta_2$ are limited to be close enough with each other to eliminate any local maximum point between them, for monotonicity.

### B. Direct roughening

A critical step of the SIR filter is resampling that generally replicates high-weighted particles to replace small-weighted ones. As a result of this, many particles may have the same state i.e. they are replications of the same particle, leading to the so-called sample impoverishment problem. To counteract this problem, one efficient solution is to spread the replicated particles by introducing additional noise, namely roughening. This can be realized in two ways that have equivalent results. One way is to increase the transition noise for particle propagation directly as called direct roughening [5], and the other is to apply roughening separately after resampling that is the separate roughening scheme proposed in [6].

In contrast to the state dynamics given in (1), the Markov transition (called propagation) of the $i$th particle that is perturbed by a roughening noise $r$ in the direct roughening approach, i.e. the proposal function, can be written as

$$x_k^{(i)} = f_{k|k-1}\left(x_{k-1}^{(i)}, \theta^* + r\right) \tag{5}$$

where $\theta^*$ is the process noise involved with the state and the roughening noise $r$ is normally zero-mean Gaussian $N(0, \Sigma r)$. In the case of sample impoverishment, the direct roughening helps to improve the approximation quality of the posterior by spreading particles in the state space. To note, over roughening (too significant $r$) however will lead to very dispersive distribution of particles and will conversely reduce the estimation accuracy. Inspired by the direct roughening approach, as long as the SIR filter suffers from sample impoverishment and will benefit from the direct roughening, we have

**Remark 2.** ∀ $\theta$ that is 'slightly' bigger than $\theta^*$: the SIR filter that uses state transition noise $\theta$ for particle propagation will obtain better approximation of the posterior than that uses the noise $\theta^*$.

**Proof.** This Remark is no more than a re-statement of the validity of the direct roughening strategy. $\theta$ is limited to be 'slightly' bigger than $\theta^*$ to eliminate any local peak between them, for monotonicity.

Combing Remark 1 and 2, it is straightforward to arrive at the assertion that the SIR filter will tend to yield a bigger-than-the-truth estimate of the state transition noise in the forward-only filtering when the filter suffers from sample impoverishment. In fact, existing experiments e.g. [1, 2, 3] have observed the bias but have not explained it adequately. In the following, further simulations are provided to demonstrate the bias of the SIR filter in estimation of the state Markov transition noise.

## III. Simulation

A sufficient condition for the occurrence of the bias of the SIR filter in estimation of the transition noise is that the filter benefits from the direct roughening approach, which will be demonstrated quantitatively below. Without loss of generality, consider estimating the static transition noise in a 1-dimensional SSM give as follows

$$x_k = 0.5x_{k-1} + \frac{25x_{k-1}}{\left(1+x_{k-1}^2\right)} + 8\cos(1.2(k-1)) + \theta \tag{6}$$

$$y_k = 0.05x_k^2 + \beta \tag{7}$$

where Gaussian noise $\theta \sim N(0, Q)$, $\beta \sim N(0, 1)$, $Q$ is the unknown variance of the zero-mean Gaussian Markov transition noise to be estimated. Without loss of generality, the state $x_k$ evolves with the transition noise with variance $Q^*=1$.

To evaluate the estimation accuracy (only in simulation), the RMSE (root mean square error) is used which is defined as follows

$$\text{RMSE} = \left( \frac{1}{T} \sum_{k=1}^{T} (x_k - \hat{x}_k)^2 \right)^{1/2} \quad (8)$$

where $\hat{x}_k$ is the estimate of the state $x_k$ which is the mean state of all particles. The RMSE is unavailable in practice since $x_k$ is unknown. The RMSD (root mean square discrepancy) between the estimated observation $\hat{y}_k$ and the real observation $y_k$ are defined as a measurement of the likelihood for offline parameter estimation

$$\text{RMSD} = \left( \frac{1}{T} \sum_{k=1}^{T} (y_k - \hat{y}_k)^2 \right)^{1/2}, \quad \hat{y}_k = g_k(\hat{x}_k) \quad (9)$$

To capture the average performance, the simulation length is T=1000 steps and each simulation runs 500 trials.

In the first simulation, four bootstrap SIR filters are designed that apply $Q$=1, 1.1, 1.2 and 1.5 respectively for particle propagation. $Q=Q^*=1$ is the basic SIR filter and $Q$=1.1, 1.2 and 1.5 are roughening-enhanced. Their average RMSE are given in Fig.1, which shows that roughening-enhanced SIR filters perform better than the basic SIR filter. Especially when the number of particles is small (e.g. between 20~60), sample impoverishment is more serious and therefore roughening is more helpful. Just as impoverishment is cases specific, the effectiveness of the roughening approach for the SIR filter depends on cases see the discussion given in [5] which is a multi-dimensional SSM. This means, the bias of the SIR filter is also cases specific.

In the second simulation, the SIR filter uses different parameters $Q$ (from 0.5 to 4 with interval 0.1) and the same 50 particles. The average RMSE and RMSD are plotted in Fig.2, which provides more details of the bias of the SIR filter in terms of estimating $Q$. As indicated the RMSE result compared with the red line, the SIR filter benefits from a state transition noise that is bigger, but not too much to prevent overshooting, than that involves with the true state. To note, RSMD is not monotonically proportional with RMSE in the whole domain but instead, the bigger the $Q$ used for particle propagation, the smaller the RMSD. Now RMSD can be used for forward-only offline estimating $Q^*$. Then, the optimal estimate of $Q$ must be a value that is bigger than the real $Q^*=1$. This directly demonstrates that the SIR filter will yield a bigger-than-the-truth estimate of (the variance of) the state transition noise, at least in the case of sample impoverishment. As stated, more experimental evidences of online parameter estimation can be found in e.g. [1, 2, 3] which are consistent with our assertion. This is also often interpreted as that the particle filter benefits from the sampling proposal with a heavy trail that is insensitive to the outliers.

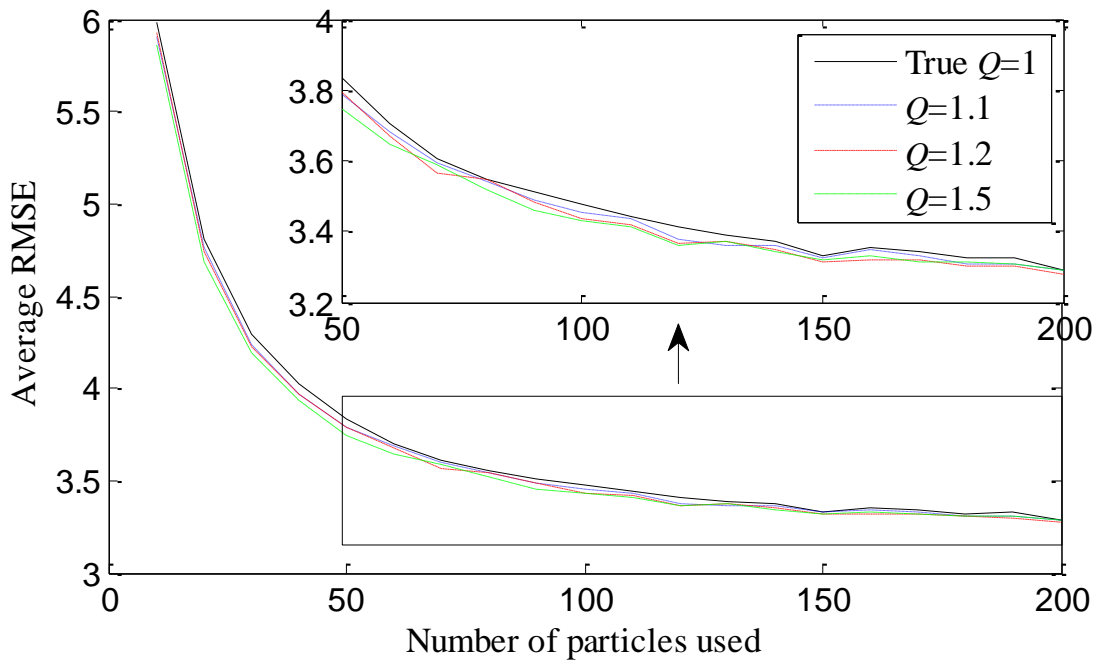

**Fig. 1** *RMSE against different number of particles used in the filter*

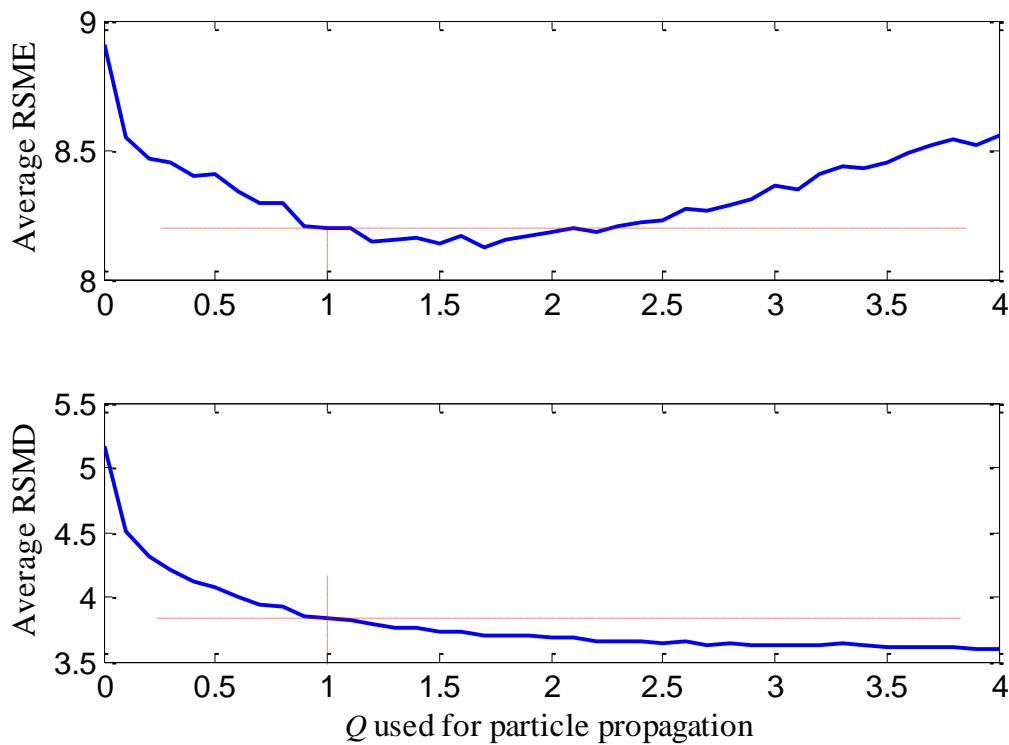

**Fig. 2** *RMSE and RMSD against different Q used in the filter*

### IV. Conclusion

A comparably big noise used for particle propagation helps to alleviate the sample impoverishment caused by resampling and will therefore produce better approximation of the posterior. As such, the SIR filter tends to yield a biased (bigger than the truth) estimate of the state Markov transition noise, especially when the forward-only filter suffers from sample impoverishment. The direct roughening approach is applied to elaborate this by means of both qualitative deduction and quantitative simulations. As sample impoverishment is customized to specific cases and is hard to be ruled out completely, so is the (degree of) bias of the SIR filter. The founding and simulation results agree with the theory which suggests that the particle filter benefits from a sampling proposal function that that has a heavier tail so that the filter is sensitive to outliers.